\documentstyle[prl,aps,multicol,epsfig]{revtex}
\tighten
\narrowtext
\begin{document}
\draft
 
\title{Quantum Billiards with Surface Scattering: 
Ballistic Sigma-Model Approach}
\author{Ya.~M.~Blanter$^{a}$,
A.~D.~Mirlin$^{b,*}$, and B.~A.~Muzykantskii$^c$} 
\address{
$^a$  D\'epartement de Physique Th\'eorique, Universit\'e de Gen\`eve,  
CH-1211 Gen\`eve 4, Switzerland\\
$^b$ Institut f\"ur Theorie der Kondensierten Materie,
Universit\"at Karlsruhe, 76128 Karlsruhe, Germany\\
$^c$ Department of Physics, University of Warwick, CV4 7AL Coventry,
UK} 
\date{December 22, 1997}
\maketitle 
\begin{abstract}
  Statistical properties of energy levels and eigenfunctions in a
  ballistic system with diffusive surface scattering are investigated.
  The two-level correlation function, the level number variance, the
  correlation function of wavefunction intensities, and the inverse
  participation ratio are calculated.
\end{abstract}

\pacs{PACS numbers: 05.45.+b, 73.23.Ps, 73.20.Dx}

\begin{multicols}{2}
  
The statistical properties of spectra of disordered diffusive systems are
now well understood. Using the supersymmetric $\sigma$-model approach it
has been possible to demonstrate the relevance of the random matrix
theory (RMT) and to calculate deviations from its predictions both for
the level \cite{Efetov83,KM,AAn} and eigenfunction \cite{FM,FE,BM}
statistics.  Generalization of these results to the case of a chaotic
{\it ballistic} system (i.e. quantum billiard) has become a topic of
great research interest.  For ballistic disordered systems the
$\sigma$-model has been proposed \cite{MK}, with the Liouville operator
replacing the diffusion operator in the action.  It has also been
conjectured that the same $\sigma$-model in the limit of vanishing
disorder adequately describes statistical properties of spectra of
individual classically chaotic system.  This conjecture was further
developed in \cite{Agam,AASA} where the $\sigma$-model was obtained by
means of energy averaging, the Liouville operator replaced by its
regularization --- the Perron-Frobenius operator and some necessary
conditions for validity of this description were put forward.

However, straightforward application of the results of
Refs.\cite{KM,AAn,FM,BM} to the case of an individual chaotic system is
complicated by the fact that the eigenvalues of the Perron-Frobenius
operator are unknown, while its eigenfunctions are extremely singular.
For this reason the $\sigma$-model approach has so far failed to provide
explicit results for any particular ballistic system.
  
To overcome this difficulty, we consider a model of a billiard with
surface disorder leading to diffusive scattering of a particle in each
collision with the boundary. This models behavior of a quantum particle
in a box with a rough boundary which is irregular on the scale of the
wave length.  Since the particle loses memory of its direction of motion
after a single collision, this model describes a limit of an ``extremely
chaotic'' ballistic system, with the typical relaxation time being of
order of the flight time. (This should be contrasted with the case of a
relatively slight distortion of an integrable billiard \cite{Roughb}.)
One might naively think that all results for such a model could be
obtained by setting $l\approx L$ in a system with bulk disorder. In fact,
the level statistics in a system with bulk disorder and arbitrary
relation between mean free path $l$ and system size $L$ were studied in
\cite{AG}.  We will see, however, that our results are qualitatively
different in some respects, which shows that systems with bulk and
surface disorder are not equivalent. On the other hand, our findings are
in agreement with general expectations for chaotic billiards based on a
semiclassical (trace formula) treatment \cite{Berry}.

To simplify the calculations, we consider a circular billiard.  A similar
problem was studied numerically in Ref. \cite{Louis} for a square
geometry. We consider only the case of unitary symmetry (broken
time-reversal invariance); generalization to the orthogonal case is
straightforward and will be given elsewhere \cite{BMM}. The level
statistics for the same problem were independently studied in
Ref. \cite{DEK}.  

{\bf Properties of the Liouville operator}. Our starting point is the
sigma-model derived recently \cite{MK} for
ballistic disordered systems. The effective action for this model has the form
\begin{eqnarray} \label{model1}
F[g(\bbox{r}, \bbox{n})] & = & \frac{\pi\nu}{4} \int d\bbox{r} {\rm Str}
\left[ i\omega \Lambda \langle g(\bbox{r}) \rangle -
\frac{1}{2\tau(\bbox{r})} 
\langle g(\bbox{r}) \rangle^2 \right. \nonumber \\
& - & \left. 2v_F \langle \Lambda U^{-1} \bbox{n} \nabla U
\rangle \right].  
\end{eqnarray}
Here $g (\bbox{r}, \bbox{n})$ is a $8\times 8$ supermatrix, which depends
on the coordinate $\bbox{r}$ and direction of the momentum $\bbox{n}$.
The angular braces denote averaging over $\bbox{n}$: $\langle {\cal O}
(\bbox{n}) \rangle = \int d{\bbox n} {\cal O}({\bf n})$ with the
normalization $\int d{\bbox n} = 1$. The matrix $g$ is constrained by the
condition $g(\bbox{r}, \bbox{n})^2 = 1$, and can be represented as $g = U
\Lambda U^{-1}$, with $\Lambda = {\rm diag} (1,1,1,1,-1,-1,-1,-1)$; see
\cite{Efetov83,AASA} for more detailed definitions.  Since we are
interested in the clean limit with no disorder in the bulk, the second
term in the action (\ref{model1}) containing the elastic mean free time
$\tau$ is zero everywhere except at the boundary where it modifies the
boundary condition (see below).

Most of the statistical properties of energy levels \cite{ASh,KM,AAn} and
eigenfunctions \cite{FM,BM} are determined by the structure of the action
in the vicinity of the homogeneous configuration of the $g$-field,
$g(\bbox{r}, \bbox{n})=\Lambda$. Writing $U = 1 - W/2+\ldots$, one finds
the action in the leading order in $W$ to be
\begin{equation} \label{model2}
  F_0[W] = -\frac{\pi\nu}{4} \int d\bbox{r} d\bbox{n} {\rm Str} \left[
    W_{21} \left( \hat K - i\omega \right) W_{12} \right],
\end{equation}
where the indices $1,2$ refer to the ``advanced-retarded'' decomposition
of $W$, and the Liouville operator $\hat K \equiv v_F \bbox{n} \nabla$.
This ``linearized'' action has the same form as that of a diffusive
system, with the diffusion operator being replaced by the Liouville
operator. This enables us to use the results derived for the diffusive
case by substituting the eigenvalues and eigenfunctions of the operator
$\hat K$ for those of the diffusion operator.

The operator $\hat K$ should be supplemented by a boundary condition,
which depends on the form of the surface roughness.  As a model
approximation we consider purely diffusive scattering \cite{Fuchs} for
which the distribution function $\varphi(\bbox{r},\bbox{n})$ of the
outgoing particles is constant and is fixed by flux conservation:
$$
\varphi(\bbox{r}, \bbox{n}) = \pi \int_{(\bbox{N}\bbox{n'}) > 0}
\left( \bbox{N} \bbox{n'} \right) \varphi (\bbox{r}, \bbox{n'})
d\bbox{n'}, \ \ \ \left (\bbox{N} \bbox{n} \right) < 0.
$$
Here the point $\bbox{r}$ lies at the surface, and $\bbox{N}$ is an
outward normal to the surface. This boundary condition should be
satisfied by the eigenfunctions of $\hat K$.

The eigenvalues $\lambda$ of the operator $\hat K$ corresponding to 
angular momentum $l$ obey the equation
\begin{equation} \label{values}
\tilde J_l(\xi) \equiv -1 + \frac{1}{2} \int_0^{\pi} d\theta \sin\theta
\exp \left[ 2il\theta + 2 \xi \sin\theta \right]  = 0,
\end{equation}
where $\xi \equiv R\lambda/v_F$, and $R$ is the radius of the circle. For
each value of $l=0,\pm1,\pm2,\ldots$ Eq.(\ref{values}) has a set of
solutions $\xi_{lk}$ with $\xi_{lk}=\xi_{-l,k}$ and
$\xi_{lk}=\xi^*_{l,-k}$. These $\xi_{lk}$ can be naturally labeled with
$k=0,\pm1,\pm2,\ldots$ (even $l$) or $k=\pm1/2,\pm3/2,\ldots$ (odd $l$).
For $l=k=0$ we have $\xi_{00}=0$, corresponding to the zero mode $\varphi
(\bbox{r}, \bbox{n})=\mbox{const}$. All other eigenvalues have positive
real part $\mbox{Re}\,\xi_{lk} > 0$ and govern the relaxation of the
corresponding classical system to the homogeneous distribution in the
phase space.

\begin{figure}
\narrowtext
{\epsfxsize=7cm\epsfysize=6.0cm\centerline{\epsfbox{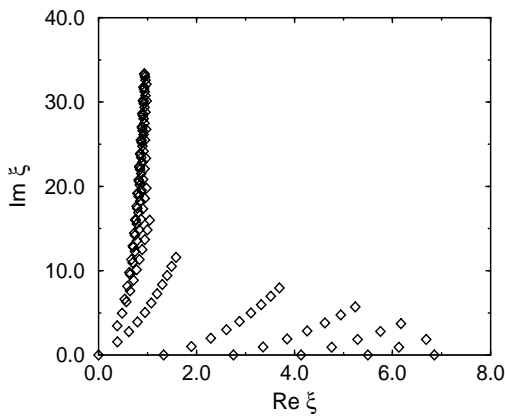}}}
\caption{First $11\times 11$  ($0 \le k,l < 11$)
eigenvalues of the Liouville operator $\hat K$ in
units of the inverse time of flight, $v_F/R$.}
\label{fig1}
\end{figure}

The asymptotic form of the solutions of Eq.(\ref{values}) for large
$\vert k \vert$ and/or $\vert l \vert$ can be obtained by using the
saddle-point method,
\begin{eqnarray} \label{valas}
\xi_{kl} \approx \left\{ \begin{array}{lr} 0.66 l + 0.14 \ln l + 0.55 \pi i
k, & 0 \le k \ll l \\ 
(\ln k)/4 + \pi i (k+1/8), & 0 \le  l \ll k
\end{array} \right. .
\end{eqnarray}
Note that for $k=0$ all eigenvalues are real, while for high values of
$k$ they lie close to the imaginary axis and do not depend on $l$ (see
Fig.~1).

{\bf Level statistics, low frequencies}. We define the level correlation
function in a standard way,
$$R_2(\omega) = (\Delta V)^{2} \langle \nu(\epsilon + \omega)
\nu(\epsilon) \rangle-1,$$
where $\nu(\epsilon)$ is the density of
states, $\Delta =(V \nu)^{-1}$ is the mean level spacing and $V = \pi
R^2$ is the area. In the range of relatively low frequencies (which for
our problem means $\omega \ll v_F/R$, see below) the function
$R_2(\omega)$ quite generally has the form \cite{KM}
\begin{eqnarray} \label{lowen}
R_2(s) & = & \delta(s) - (\pi s)^{-2} \sin^2 \pi s \nonumber \\
& + & A \left( R \Delta /\pi v_F \right)^2  \sin^2 \pi s,
\end{eqnarray}
where $s = \omega/\Delta$. The first two terms correspond to the
zero-mode approximation and are given by RMT, while the last one
represents the non-universal correction to the RMT results. It contains
information about the operator $\hat K$ through the dimensionless
constant $A=\sum'\xi_{kl}^{-2}$, where the prime indicates that the
eigenvalue $\xi_{00}=0$ is excluded.  The value of $A$, as well as the
high-frequency behavior of the level correlator (see below), can be found
from the Altshuler-Shklovskii type spectral function \cite{ASh}:
\begin{equation} \label{sum1}
S (\omega) = \sum_{l} S_l (\omega);\ \ \ S_l(\omega) \equiv \sum_k 
\left( \lambda_{kl} - i\omega \right)^{-2}.
\end{equation}
According to the Cauchy theorem, $S_l$ can be
represented as an integral in the complex plane,
$$S_l (\omega) = \left( \frac{R}{v_F} \right)^2 \frac{1}{2\pi i} \oint_C
\frac{1}{(z - i\omega R/v_F)^2} \frac{\tilde J_l' (z)}{\tilde J_l (z)}
dz,$$
where the contour $C$ encloses all zeroes of the function $\tilde
J_l (z)$. Evaluating the residue at $z=i\omega R/v_F$, we find
\begin{equation}
\label{sl}
S_l (\omega) = - (R/v_F)^2 \left. \frac{d^2}{d z^2} \right\vert_{z =
  i\omega R/v_F} \ln \tilde J_l (z) .
\end{equation}
Considering the limit $\omega \to 0$ and subtracting the
contribution of $\lambda_{00}=0$, we get
\begin{equation} \label{s0}
  A = -19/27 - 175 \pi^2/1152 + 64/(9\pi^2) \approx -1.48.
\end{equation}
In contrast to the diffusive case, this constant is negative: the level
repulsion is enhanced with respect to result for RMT.  Eq.(\ref{lowen})
is valid as long as the correction is small compared to the RMT result,
i.e. provided $\omega$ is below the inverse time of flight, $v_F/R$,
which plays the role of the Thouless energy for our problem.

{\bf Level statistics, high frequencies}.  In the range $\omega\gg\Delta$
the level correlation function can be decomposed into the smooth
Altshuler-Shklovskii part $R_2^{\mbox{\small AS}} (\omega) = (\Delta^2/
2\pi^2) {\rm Re}\,S(\omega)$ \cite{ASh} and the part $R_2^{\mbox{\small osc}}$
which oscillates on the scale of the level spacing.  Evaluating the asymptotic
behavior of $S_l(\omega)$ from Eq.(\ref{sl}), we find in the
high-frequency regime when $\omega \gg v_F/R$:
\begin{equation} \label{highen0}
R_2^{\mbox{\small AS}} (\omega) = 
\left( \frac{\Delta R}{v_F} \right)^2 \left(
\frac{v_F}{2\pi\omega R} \right)^{1/2} \cos \left( 4 \frac{\omega R}{v_F} -
\frac{\pi}{4} \right).
\end{equation} 

The oscillating part of the level correlation function
$R_2^{\mbox{osc}}(s)$ for frequencies $\omega \gg
\Delta$ is given by \cite{Agam}
\begin{equation} \label{highen}
R_2^{\mbox{osc}} (s)  = (1/2\pi^2)\cos(2\pi s) D(s), 
\end{equation}
where $D(s)$ is the spectral determinant,
$$D(s) = s^{-2}\prod_{kl\ne(00)}  (1- is\Delta/\lambda_{kl} )^{-1} 
(1 + is\Delta/\lambda_{kl} )^{-1}.$$ 
Since $\Delta^{-2}\partial^2\ln D(s)/\partial
s^2=-2\mbox{Re}\,S(\omega)$, we can restore $D(s)$ from
Eqs.(\ref{sum1}), (\ref{sl}), up to a factor of the form $\exp(c_1 +
c_2 s)$, with $c_1$ and $c_2$ being arbitrary constants. These
constants are fixed by the requirement that Eq.(\ref{highen}) in the
range $\Delta\ll\omega\ll v_F/R$ should reproduce the low-frequency
behavior (\ref{lowen}).  As a result, we obtain
\begin{equation} \label{spdet1}
D(s) = \left( \frac{\pi}{2} \right)^6 \frac{1}{N} \prod_l
\frac{1}{\tilde J_l (i s N^{-1/2}) \tilde J_l (-i s N^{-1/2})}.
\end{equation}
For high frequencies $\omega \gg v_F/R$ this yields the following
expression for the oscillating part of the
level correlation function:
\begin{equation} \label{highen2}
R_2^{\mbox{osc}} (\omega) = \frac{\pi^4}{128} 
\left( \frac{\Delta R}{v_F} \right)^2
\cos \left( \frac{2\pi \omega}{\Delta} \right).
\end{equation} 
It is remarkable that the amplitude of the oscillating part does not
depend on frequency. This is in contrast to the diffusive case, where in
the Altshuler-Shklovskii regime ($\omega$ above the Thouless energy) the
oscillating part $R_2^{\mbox{osc}} (\omega)$ is exponentially small
\cite{AAn}.

{\bf The level number variance}. The smooth part of the level
correlation function can be best illustrated by plotting the variance
of the number of levels in an energy interval of width $E = s\Delta$, 
\begin{equation} \label{lnv}
\Sigma_2 (s) = \int_{-s}^s \left(s - \vert \tilde{s} \vert
\right) R_2(\tilde{s}) d\tilde{s},  
\end{equation}
A direct calculation gives the following asymptotic behavior: 
\begin{equation} 
  \pi^2 \Sigma_2 (s) 
  =  1 + \gamma + \ln (2\pi s) +
    A s^2/(2N) \label{lnv1}
\end{equation}
when $ s \ll N^{1/2}$ and
\begin{eqnarray}
  \pi^2 \Sigma_2 (s) &=& 1 + \gamma + \ln \frac{16 N^{1/2}}{\pi^2}
  \nonumber \\ 
  &&  - \frac{\pi^2}{16}  \left( \frac{2N^{1/2}}{\pi s}
  \right)^{1/2} \cos \left( \frac{4 s}{N^{1/2}} - \frac{\pi}{4} \right)
\label{lnv2}
\end{eqnarray}
when $s \gg N^{1/2}$.  Here $N =(v_F/R\Delta)^2 = (p_F R/2)^2$ is the
number of electrons below the Fermi level, $\gamma \approx 0.577$ is
Euler's constant, and $A$ is defined by Eq.(\ref{s0}). The first three
terms at the rhs of Eq.(\ref{lnv1}) represent the RMT contribution (curve
1 in Fig.~2).  The two asymptotics (\ref{lnv1}) and (\ref{lnv2}) are
shown in Fig.~1 as curves 2 and 3 respectively.

\begin{figure}
\narrowtext
{\epsfxsize=7cm\epsfysize=5.0cm\centerline{\epsfbox{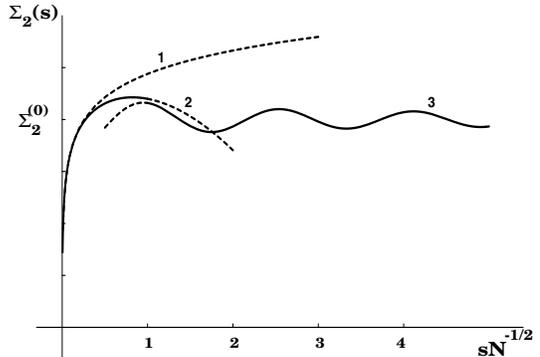}}}
\caption{Level number variance $\Sigma_2 (E)$ as a function of
energy; $s = E/\Delta$. Curve 1 shows the RMT result, while curves 2
and 3 correspond to asymptotic regimes of low (\protect\ref{lnv1}) and
high (\protect\ref{lnv2}) frequencies. The saturation value
$\Sigma_2^{(0)}$ is given in the text.} 
\label{fig2}
\end{figure}

As seen from Fig.2, the two curves (\ref{lnv1}) and (\ref{lnv2})
perfectly match in the intermediate regime, $s\sim N^{1/2}$, and taken
together they provide a complete description of $\Sigma_2(s)$.  According
to Eq.(\ref{lnv2}), the level number variance saturates at the value
$\Sigma_2^{(0)} = \pi^{-2} (1 + \gamma + \ln (16N^{1/2}/\pi^2))$, in
contrast to the behavior found for diffusive systems \cite{ASh} or
ballistic systems with weak bulk disorder \cite{AG}. The saturation
occurs at energies $s \sim N^{1/2}$, or in conventional units $E \sim
v_F/R$.  This behavior of $\Sigma_2 (s)$ is expected for a generic
chaotic billiard \cite{Berry}. It is also in good agreement with the
results for $\Sigma_2 (s)$ found numerically for a tight-binding model
with moderately strong disorder on boundary sites \cite{Louis}.

{\bf Eigenfunction statistics}. Now we study correlations of
the amplitudes of an eigenfunction in two different points. Correlation of
different eigenfunctions will be considered elsewhere \cite{BMM}. 
Following Ref. \cite{BM}, we define
$$
\alpha (\bbox{r_1}, \bbox{r_2}, E) = \Delta V^2 \langle
\sum_{\mu} \vert \psi_{\mu} (\bbox{r_1}) \psi_{\mu} (\bbox{r_2})
\vert^2 \delta (E - \epsilon_{\mu}) \rangle,
$$
where $\psi_{\mu}$ are the eigenfunctions corresponding to the exact
single-particle states $\mu$. A calculation analogous to that of
Ref.\cite{BM} yields
\begin{equation} \label{corfun1}
\alpha (\bbox{r_1}, \bbox{r_2}, E) = 1 + \Pi (\bbox{r_1}, \bbox{r_2}),
\end{equation}
where $\Pi$ is the Green's function of the operator $\hat K$
integrated over directions of momentum, $\Pi (\bbox{r_1}, \bbox{r_2})
= \int d\bbox{n_1} d\bbox{n_2}\, g(\bbox{r_1}, \bbox{n_1}; \bbox{r_2},
\bbox{n_2})$. Here $g$ is the full Green's function of the operator
$\hat K$, i.e. a solution to the equation
\begin{eqnarray} \label{green0}
& & \hat K g(\bbox{r_1}, \bbox{n_1}; \bbox{r_2}, \bbox{n_2}) =
\nonumber \\
& & \left( \pi \nu \right)^{-1} \left[ \delta(\bbox{r_1} - \bbox{r_2})
\delta(\bbox{n_1} - \bbox{n_2}) - V^{-1} \right].
\end{eqnarray} 
Direct calculation gives:
\begin{eqnarray} \label{green}
\Pi (\bbox{r_1}, \bbox{r_2}) & = & \Pi_1 (\bbox{r_1}, \bbox{r_2}) +
\Pi_2 (\bbox{r_1}, \bbox{r_2}), \nonumber \\
\Pi_1 (\bbox{r_1}, \bbox{r_2}) &=& \tilde{k}_d(\bbox{r_1} - \bbox{r_2})-
V^{-1}\int d\bbox{r'_1}\tilde{k}_d(\bbox{r'_1} - \bbox{r_2}) \\ 
&&\hspace{-2cm}
-V^{-1}\int d\bbox{r'_2}\tilde{k}_d(\bbox{r_1} - \bbox{r'_2})
+V^{-2}\int d\bbox{r'_1}d\bbox{r'_2}
\tilde{k}_d(\bbox{r'_1} - \bbox{r'_2}); \nonumber\\
\Pi_2 (\bbox{r_1}, \bbox{r_2}) & = & \frac{1}{4\pi p_FR}
\sum_{k=1}^{\infty} \frac{4k^2 - 1}{4k^2} \left( \frac{r_1r_2}{R^2}
\right)^k \cos k \left( \theta_1 - \theta_2 \right) \nonumber
\end{eqnarray}
where $\tilde{k}_d(\bbox{r}) = 1/(\pi p_F|\bbox{r}|)$, and
$(r,\theta)$ are the polar coordinates. This formula has
a clear interpretation. The function $\Pi$ can be represented as a sum
over all possible paths leading from $\bbox{r_1}$ to $\bbox{r_2}$, with
possible surface scattering in between. In particular, the function
$\Pi_1$ corresponds to trajectories coming directly from $\bbox{r_1}$ to
$\bbox{r_2}$ with no reflection from the surface. Therefore, the term
$\Pi_1$ is universal and is not sensitive to the geometry of the system.
It can be obtained from the RMT-like prediction that amplitudes of
different wavefunctions are independent Gaussian variables \cite{Berry1}.
More precisely, we find that the function $k_d({\bf r_1}-{\bf
r_2})=J_0^2(p_F|{\bf r_1}-{\bf r_2}|)$ of Refs.\cite{BM,Berry1} is
replaced in Eq.(\ref{green}) by its smoothed version, $\tilde{k}_d({\bf
r_1}-{\bf r_2})=1/(\pi p_F|{\bf r_1}-{\bf r_2}|)$. This is because our
semiclassical approach is valid on scales much larger than the wave
length. To cure this flaw, one has to replace $\tilde{k}_d({\bf r})$ by
$k_d({\bf r})$ in the expression Eq.(\ref{green}) for $\Pi_1$.  The
second term, $\Pi_2$, is due to the surface scattering. It can be shown
\cite{BMM} that in the numerator $4k^2-1$ the first term comes from
trajectories with only one surface reflection, while the second sums up
contributions from multiple reflections. A formula analogous to
(\ref{corfun1}) was proposed very recently for a generic chaotic system
\cite{Sred}.

Finally, we calculate the inverse participation ratio, $\langle
P_2\rangle \equiv V^{-2} \int d\bbox{r} \alpha(\bbox{r}, \bbox{r}),$
which characterizes the degree of spatial uniformity of eigenfunctions.
The RMT prediction for this quantity, $P_2^{(0)}=2/V$ is recovered from
Eqs.(\ref{corfun1}), (\ref{green}) if we take into account the first term
in the expression for $\Pi_1$, since $k_d(0)=1$. The leading correction
comes from the single-reflection contribution to the term $\Pi_2$, and is
equal to
$$\delta P_2 = V^{-1} (4\pi p_F R)^{-1} \ln (p_F R) \sim 
P_2^{(0)} N^{-1/2} \ln N.$$

In conclusion, we have used the ballistic $\sigma$-model approach to
study statistical properties of levels and eigenfunctions in a billiard
with diffusive surface scattering, which exemplifies a ballistic system
in the regime of strong chaos.  We have found that the level repulsion
and the spectral rigidity are enhanced compared to RMT. In particular,
the level number variance saturates at the scale of the inverse time of
flight, in agreement with Berry's prediction for a generic chaotic system
\cite{Berry}.  As another manifestation of the strong spectral rigidity,
the oscillating part of the level correlation function does not vanish at
large level separation. We have also considered correlations of eigenfunction
amplitudes in different spatial points and calculated the deviation of
the inverse participation ratio from the RMT value.

This work was supported by the Swiss National Science Foundation and
SFB 195 der Deutschen Forschungsgemeinschaft. We are grateful to the
University of Geneva (A.~D.~M.) and Isaac Newton Institute for
Mathematical Sciences of the Cambridge University, where part of this
work was done, for hospitality and support.

\end{multicols}

\end{document}